\documentclass[a4paper]{jpconf}

\usepackage{graphicx}
\usepackage[latin1]{inputenc}
\usepackage[spanish,english]{babel}

\newcommand{\be}{\begin{equation}}
\newcommand{\ee}{\end{equation}}
\newcommand{\bea}{\begin{eqnarray}}
\newcommand{\eea}{\end{eqnarray}}

\begin{document}
\title{Inflation, Renormalization, and CMB Anisotropies}

\author{I. Agulló$^1$, J. Navarro-Salas$^2$, Gonzalo J. Olmo$^{3,1}$, and Leonard Parker$^1$}
\address{ {\footnotesize 1. Physics Department, University of
Wisconsin-Milwaukee, P.O.Box 413, Milwaukee, WI 53201 \  USA }\\
{\footnotesize 2. Departamento de Física Teórica and
IFIC, Centro Mixto Universidad de Valencia-CSIC.
    Facultad de Física, Universidad de Valencia,
        Burjassot-46100, Valencia, Spain. } \\
{\footnotesize 3. Instituto de Estructura de la Materia, CSIC, Serrano 121, 28006 Madrid, Spain}
}
\ead{ivan.agullo@uv.es, jnavarro@ific.uv.es, olmo@iem.cfmac.csic.es, leonard@uwm.edu}


\begin{abstract}
In single-field, slow-roll inflationary models, scalar and tensorial (Gaussian) perturbations are both characterized by a zero mean and a non-zero variance. 
In position space, the corresponding variance of those fields diverges in the ultraviolet. The requirement of a finite variance in position space forces its regularization via quantum field renormalization in an expanding universe. This has an important impact on the predicted scalar and tensorial power spectra for wavelengths that today are at observable scales. In particular, we find a non-trivial  change in the consistency condition that relates the tensor-to-scalar ratio $r$ to the spectral indices. For instance, an exact scale-invariant tensorial power spectrum, $n_t=0$, is now compatible with a non-zero ratio $r\approx 0.12\pm0.06$, which is forbidden by the standard prediction ($r=-8n_t$). Forthcoming observations of the influence of relic gravitational waves on the CMB will offer a non-trivial test of the new predictions.
\end{abstract}

\section{Introduction}
	
Inflation has become a fundamental piece of the standard cosmological model \cite{books}. Not only it addresses and solves the classical problems of the big bang model but it also provides an elegant quantum mechanical mechanism for the origin of primordial perturbations, essential for explaining the structures that we see today \cite{inflation2}. In the simplest models of inflation, a scalar field slowly rolling down its potential causes an exponentially large expansion. In this background, quantum vacuum fluctuations of the inflaton field itself and of purely tensorial modes (gravitational waves) acquire classical properties due to their interaction with the expanding geometry. As a result, a classical field of scalar (and tensorial) inhomogeneities arises. Cosmological perturbation theory tells us that these primordial perturbations are responsible for the existence and richness of the structure that we observe in the temperature anisotropies of the cosmic microwave background (CMB) and in the large scale distribution of galaxies. \\
A disturbing aspect of the spectra of scalar and tensorial perturbations generated during inflation is that they have a divergent variance. In the case of classical perturbations, divergences of the variance are usually removed by means of window functions that filter out the wavelengths that cause the problem. However, the primordial spectra have a quantum origin and their divergences should be removed, on grounds of theoretical consistency, by means of the well-established methods of renormalization in curved spaces\cite{parker-toms}. In this work we show that the regularization of the variance of scalar and tensorial perturbations during inflation has a non-trivial impact on the primordial power spectra of those fields \cite{Parker07,collaboration}.  Though the resulting spectra are almost scale free within the slow-roll approximation, the relation between spectral indices and the slow-roll parameters is significantly changed as compared to the standard derivation. The ratio of tensor to scalar amplitudes is also modified and gives rise to a number of possibilities which are explicitly forbidden according to the standard predictions. In particular, the tensorial spectral index is allowed to take a zero value without implying a zero tensor to scalar ratio, or may even take positive values.

\section{Spectrum of fluctuations from inflation}

Let us first focus on the production of relic gravitational waves by considering fluctuating tensorial modes $h_{ij}(\vec{x},t)$ in an expanding, spatially flat universe
\be ds^2= -dt^2 + a^2(t)(\delta_{ij} + h_{ij})dx^idx^j \ . \ee
The wave equation obeyed by these modes is given by
\be -a^2 \ddot h_{ij} -3a\dot a \dot h_{ij} + \nabla^2 h_{ij} =0  \ . \label{eq:hij} \ee
The fluctuating fields $h_{ij}$ can be decomposed in two independent polarization modes $h^{(p)}(\vec{x},t) e^{(p)}_{ij}$,
where the index $p$ represents the two polarizations $+$ and $\times$, and  $e^{(p)}_{ij}$ are two independent constant matrices that  obey the conditions $e_{ij}=e_{ji}$,  $\sum_ie_{ii}=0$, and $\sum_ik_ie_{ij}=0$. Expanding the fields  $h^{(p)}(\vec{x},t)$ in modes and omitting the polarization label $p$,  (\ref{eq:hij}) yields 
\be \label{waveq}
\ddot{h}_k + 3H \dot{h}_k +\frac{k^2}{a^2} h_k
=0 \ , \ee
with $k\equiv |\vec{k}|$. This equation indicates that the relevant part of the tensorial modes is represented by two independent massless scalar fields $h^{(+,\times)}(\vec{x},t)$. The solutions that satisfy the asymptotic adiabatic condition \cite{parker69,parker-toms} for very high $k$ in a slow-roll inflationary background are 
\be \label{amplten} 
h_{k}(t) = (-16\pi G\tau \pi/4(2\pi)^3a^2)^{1/2}H^{(1)}_{\nu}(-k\tau) \ , 
\ee
where $\tau=\int dt/a(t)$ is the conformal time, the index of the Bessel function is  $\nu=3/2 + \epsilon$, and $\epsilon=\frac{M_P^2}{2}\left(\frac{V'}{V}\right)^2$ is one of the slow-roll parameters. 
With these solutions it is easy to evaluate the variance of the gravitational wave fields $h^{(+,\times)}$
 \be \label{eq:varh-unren}
\langle h^2 \rangle =\int_0^{\infty}k^2dk \int d\Omega |h_k|^2=
\int_0^{\infty}\frac{dk}{k}\Delta^2_h (k,t) \ ,
\ee
where we have defined the power spectrum of the field $h$ as $\Delta^2_h (k,t)\equiv 4\pi k^3 |h_k|^2$. As we advanced in the introduction, the large $k$ behavior of the modes makes this integral divergent
\be \label{divergentintegral}\langle h^2 \rangle =
\int_0^{\infty}\frac{dk}{k} \frac{16\pi G k^3}{4\pi^2a^3}\left [
\frac{a}{k}[1 +\frac{(2+3\epsilon)}{2k^2\tau^2}] + ... \right] \ .
\ee
Since the power spectrum is expressed
in momentum space, the natural renormalization scheme to apply is
the so-called adiabatic subtraction \cite{Parker07}, as it
renormalizes the theory in momentum space. Adiabatic renormalization
\cite{ parker-toms, birrel-davies} removes the
divergences present in the formal expression (\ref{eq:varh-unren}) by
subtracting counterterms mode by mode in the integrand of (\ref{eq:varh-unren}) 
\be
\label{eq:varh-ren} \langle h^2 \rangle_{ren} = \int_0^{\infty}\frac{dk}{k}\tilde{\Delta}^2_h (k,t) =
\int_0^{\infty}\frac{dk}{k}\left[4\pi k^3|h_{\vec{k}}|^2 - \frac{16\pi G
k^3}{4\pi^2 a^3}(w_k^{-1}+ (W_k^{-1})^{(2)})\right]\ ,\ee with $w_k=k/a(t)$. The
subtraction of the first term $(16\pi G k^3/4\pi^2 a^3 w_k)$
cancels the typical  flat space vacuum fluctuations, which are responsible for the quadratic divergence in the integral (\ref{divergentintegral}).
The additional term, proportional to $(W_k^{-1})^{(2)}=-\frac{1}{w_k^2}\left[\frac{1}{2}w_k^{-1/2}\frac{d^2}{dt^2}w_k^{-1/2} - \frac{1}{2}w_k^{-1}\left(\frac{3}{4}\frac{\dot{a}^2}{a^2}+\frac{3}{2}\frac{\ddot{a}}{a}\right)\right]$   and which  involves
$\dot{a}^2$ and $\ddot{a}$, is necessary to properly perform the
renormalization in an expanding universe, since it cancels the logarithmic divergence in (\ref{divergentintegral}). \\
According to expression (\ref{eq:varh-unren}), the unrenormalized power spectrum of tensor perturbations evaluated a few e-folds after the Hubble horizon exit time $t_k$ (defined by $k/a(t_k)=H(t_k)$) is given by
\begin{equation}
P_t(k)=4\Delta^2_h(k)=16\pi k^3|h_k|^2=\frac{8}{M_P^2}\left(\frac{H(t_k)}{2\pi}\right)^2
\end{equation}
The $k$ dependence of $H^2(t_k)$, $d \ln H(t_k)/d\ln k= -\epsilon $, leads to $P_t(k,t_k) =  P_t(k_0)\left(\frac{k}{k_0}\right)^{-2\epsilon}$, 
where $k_0$ is a pivot scale. The tensorial spectral index $n_t$ is defined as the exponent in that expression. So $n_t=-2\epsilon (t_k)$. Using the renormalized variance (\ref{eq:varh-ren}) to define the renormalized power spectrum and taking into account the time dependence of the counterterms, we find 
\be 
P_t^{ren}(k,n)=4\tilde\Delta_h^2(k,n)\approx \frac{8}{M_P^2}\left (\frac{H(t_k)}{2\pi} \right )^2 \epsilon(t_k) (2n-3/2)\ , \label{eq:Dh-n2}\ee
where $n$ represents the number of $e$-folds after the Hubble exit at which the spectrum is evaluated (we assume $n>1$ but $n\epsilon \ll 1$). We note that if one evaluates the power spectra at the end of the slow-roll era (where $n\epsilon \sim 1$) the contribution of the counterterms is still significant. However, we find it more natural to evaluate the spectra soon after $t_k$, when the modes have already acquired classical properties. The spectral index corresponding to the renormalized tensorial power spectrum is given by
\begin{equation}
 n^{ren}_t\equiv \frac{d\ln P^{ren}_t}{d\ln k}= 2(\epsilon-\eta)
\end{equation}
where the slow-roll parameters $\epsilon$ and $\eta=M_P^2(V''/V)$ are evaluated at $t_k$. \\
The discussion just presented regarding the tensorial modes can be paralleled to the case of scalar perturbations. In this case one deals with the gauge invariant scalar ${\cal{R}}= \Psi+\frac{H}{\dot\phi_0}\delta \phi$, where $\Psi$ is the curvature perturbation ($R^{(3)}=4\nabla^2\Psi /a^2$) of  the spatial metric $g_{ij}=a^2[(1-2\Psi)\delta_{ij}+2\partial_{ij}E]$. In momentum space, $\cal{R}$ obeys  the
 equation \cite{books}\be \frac{d^2 {\cal{R}}_k}{d{\tau}^2} +
\frac{2}{z}\frac{dz}{d\tau}\frac{d {\cal{R}}_k}{d{\tau}} +
k^2{\cal{R}}_k =0 \ , \ee where $z\equiv a\dot{\phi}_0/H$. In the slow roll approximation $z^{-1}dz/d\tau= a H(1+2\epsilon-\eta)$, and the solutions obeying the adiabatic condition (and the de Sitter symmetry for $H$ constant) are
\be {\cal{R}}_k( t) = (-\pi \tau /4(2\pi)^3 z^2)^{1/2}H^{(1)}_{\mu}(-\tau k) \ , \ee  where $\mu= 3/2 + 3\epsilon -\eta$. The resulting unrenormalized and renormalized scalar power spectra and spectral indices are ($P_{\cal R}(k) = P_{\cal R}(k_0)\left(\frac{k}{k_0}\right)^{n_s-1} $)
\begin{eqnarray}
P_{\cal R}(k) = \frac{1}{2M_P^2\epsilon}\left(\frac{H(t_k)}{2\pi}\right)^2 & , & n_s-1=-6\epsilon +2\eta \\
P^{ren}_{\cal R}(k,n)\approx \frac{(3\epsilon - \eta)}{2M_P^2\epsilon}\left
(\frac{H(t_k)}{2\pi}\right)^2 (2n-3/2) & , & n_s-1=-6 \epsilon+2 \eta+\frac{ (12 \epsilon^2-8\epsilon \eta+\xi) }{3 \epsilon-\eta}
\end{eqnarray}
where $\xi$ is another slow roll parameter: $\xi\equiv M_P^4(V'V'''/V^2)$. This parameter can be reexpressed in terms of $\epsilon, \eta$ and the running of the tensorial index $n'_t\equiv dn_t/d\ln k$ as $n'_t= 8\epsilon (\epsilon - \eta)+2\xi$.

\section{Conclusions}

The expressions derived above for the power spectra and spectral indices show that renormalization has a significant impact on the predictions of slow-roll inflation. Unlike in the standard derivation, we see that the renormalized tensorial power spectrum is no longer directly related to the Hubble scale during inflation. The amplitude of this spectrum is now modulated by the slow-roll parameter $\epsilon$, which is zero in exact de Sitter inflation. On the other hand, the appearance of the combination $3\epsilon-\eta$ in the numerator of the renormalized scalar spectrum suggests that $P^{ren}_{\cal R}(k,n)\sim \frac{1}{2M_P^2}\left (\frac{H(t_k)}{2\pi}\right)^2$ (up to the numerical factor $2n-3/2$), which indicates that $H(t_k)\sim 10^{15}$GeV. For the exponential potential \cite{books}, we find that $H\sim 1.5\times 10^{15}/\sqrt{4n-3}$ GeV, roughly up to an order of magnitude larger than the unrenormalized prediction. \\
The most dramatic difference between the standard predictions and the predictions after renormalization presented here affects the so-called consistency condition. The tensor to scalar ratio in the standard approach gives $r=P_t/P_{\cal R}=16\epsilon=-8n_t$. This implies that if a background of gravitational waves is observed, then its spectral index must be constrained by the ratio of the tensor to scalar amplitudes. If this constraint is not satisfied, then single field slow-roll inflation would be ruled out. However, the consistency condition after renormalization becomes
\be \label{eq:r} r=4(1-n_s-n_t)+\frac{4n'_t}{n_t^2-2n'_t} \left(1-n_s-\sqrt{2 n'_t+(1-n_s)^2-n_t^2} \right) \ . \ee
This expression is much more involved that the standard one and its implications are far reaching. It allows for a zero $n_t$ while having a non-zero ratio $r$. Also, the values of $n_t$ are not constrained to be negative. In addition, according to the standard derivation, the running of $n_t$ is fully determined by the values of $n_s$ and $n_t$, since then one finds $n'_t=-n_t(1-n_s + n_t)$. On the contrary, the manipulations that lead to (\ref{eq:r}) indicate that $n'_t$ is now an independent quantity that needs to be measured in order to check the new consistency relation (\ref{eq:r}). This aspect could make more challenging the experimental verification of the consistency condition of (single-field) slow roll inflation. \\
To conclude, the predictions presented in this talk will soon come within the range of ongoing and planned experiments. If single field slow-roll inflation is correct, then observations will (hopefully) tell us if nature have chosen the renormalized spectra as the seeds of perturbations or not. Quantum field theory in curved spacetimes will thus face a crucial experimental test. \\
\noindent { \bf Acknowledgements.} This
work has been partially supported by the spanish grant FIS2008-06078-C03-02. I.A. and L.P.  have been partly
supported by NSF grants PHY-0071044 and PHY-0503366 and by a UWM RGI grant.  G.O.
thanks MICINN for a JdC contract and the ``Jos\'e Castillejo'' program for funding a stay at the University of Wisconsin-Milwaukee.

\end{document}